\documentclass{article}
\pdfoutput=1
\usepackage{etoolbox}
\makeatletter
\patchcmd{\maketitle}{\@copyrightspace}{}{}{}
\let\proof\@undefined
\let\endproof\@undefined
\makeatother

\usepackage{enumerate}
\usepackage{url}
\usepackage{ulem}
\usepackage{subfigure}
\usepackage{a4wide}
\usepackage{color}
\usepackage{enumitem}

\usepackage{amsmath}
\usepackage{amsfonts}
\usepackage{amsthm}
\usepackage{graphicx}
\usepackage{authblk}

\usepackage{epstopdf}
\usepackage{color}
\def\etal{\emph{et al.}}

\begin{document}
\title{Sic Transit Gloria Manuscriptum:\\ Two Views of the Aggregate Fate of Ancient Papers}
\author[1]{Mayank Singh}
\author[2]{Rajdeep Sarkar}
\author[1]{Pawan Goyal}
\author[1]{Animesh Mukherjee}
\author[3]{Soumen Chakrabarti}
\affil[1,3]{Department of Computer Science and Engineering}
\affil[2]{Department of Mathematics}
\affil[1,2]{Indian Institute of Technology, Kharagpur, India}
\affil[3]{Indian Institute of Technology, Bombay, India}
\affil[ ]{mayank.singh@cse.iitkgp.ernet.in, rajdeep.sarkar@iitkgp.ac.in}
\affil[ ]{\{pawang,animeshm\}@cse.iitkgp.ernet.in, soumen.chakrabarti@gmail.com}
\renewcommand\Authands{ and }

\maketitle

\begin{abstract}
When PageRank began to be used for ranking in Web search, a concern soon arose that older pages have an inherent --- and potentially unfair --- advantage over emerging pages of high quality, because they have had more time to acquire hyperlink citations. Algorithms were then proposed to compensate for this effect.  Curiously, in bibliometry, the opposite concern has often been raised: that a growing body of recent papers crowds out older papers, resulting in a collective amnesia in research communities, which potentially leads to reinventions, redundancies, and missed opportunities to connect ideas. A recent paper by Verstak \etal\ reported experiments on Google Scholar data, which seemed to refute the amnesia, or aging, hypothesis.  They claimed that more recently written papers have a larger fraction of outbound citations targeting papers that are older by a fixed number of years, indicating that ancient papers are alive and well-loved and increasingly easily found, thanks in part to Google Scholar.  In this paper we show that the full picture is considerably more nuanced. Specifically, the fate of a \emph{fixed sample} of papers, as they age, is rather different from what Verstak \etal's study suggests: there is clear and steady abandonment in favor of citations to newer papers. The two apparently contradictory views are reconciled by the realization that, as time passes, the number of papers older than a fixed number of years grows rapidly. 
\end{abstract}

\section{Introduction} \label{section:Introduction}

The volume of scholarly publication per year has grown dramatically in the last two decades, along with an explosion in the number of papers that are available online, indexed, and searchable.  The effect of such enhanced access on subsequent research is not entirely clear: there have been recent conflicting claims.  Parolo \emph{et al.} \cite{2015arXiv150301881D} present evidence that it is becoming ``increasingly difficult for researchers to keep track of all the publications relevant to their work''.  Based on analysis of citation data, they propose a pattern of a paper's citation counts per year, which peaks within a few years and then the typical paper fades into obscurity.  This work has seen considerable press following, with headlines ranging from the tongue-in-cheek ``Study shows there are too many studies'' to the more alarmist ``Science is `in decay' because there are too many studies''.  Chakraborty \emph{et al.} \cite{Chakraborty:2015:CSC:2817191.2701412} present a more nuanced analysis that clusters papers into the ephemeral and the enduring, giving some hope that not all creativity is lost in the sands of time.  Meanwhile, Verstak \emph{et al.} \cite{DBLP:journals/corr/VerstakASHILS14}, from the Google Scholar team, claim that fear of evanescence is misplaced, and that older papers account for an increasing fraction of citations as time passes. Superficially, these claims seem to be at odds with each other. But their simultaneous validity is easily explained by mild additional probing of citation statistics, which we present in this paper.  Our main observations are as follows.
\begin{itemize}[noitemsep]
\item We confirm that the fraction of citations made in a paper $p_0$ to other papers that are older than $p_0$ by a fixed time window (say, 10 to 15 years older) grows as we sample $p_0$ from more and more recent years.
\item But this happens because the number of papers becoming older than a threshold is steeply increasing with time.
\item If instead we fix a (sampled) set of papers and trace the rate of citations to them as time progresses, aging and passing into obscurity are strongly visible. 
\item These observations for citing behavior hold true across many disciplines (computer science, biomedical), sub-disciplines (algorithms, theory) as well as conferences.
\end{itemize}

\section{Dataset}
We begin our analysis using a dataset from the computer science domain, crawled from Microsoft Academic Search (MAS)\footnote{http://academic.research.microsoft.com}.
The dataset consists of bibliographic information of papers, the title of the paper, a unique index 
for the paper, its author(s), the affiliation of the author(s),
the year of publication, the publication venue, references, citation contexts, the related field(s) of the paper, the abstract and the keywords of the papers \cite{Chakraborty:2014}.  In addition, we also use another dataset available from the biomedical domain\footnote{http://www.ncbi.nlm.nih.gov/pmc/tools/ftp, downloaded in May, 2014}.  Some general statistics of both data sets are shown in Table \ref{tab:dataset}.

\begin{table}[!thb]
 \centering
 \caption{General statistics about the Computer Science (CS) and Biomedical (BM) dataset.}\label{dataset}
  \begin{tabular}{|c|c|c|}
  \hline
  &CS&BM\\\hline
  Sub-fields&25&1\\\hline
  Publication count&1,359,338&801,252\\\hline
  Author count&138,923&1,985,890\\\hline
  Year range&1960-2010&1996-2014\\\hline
 \end{tabular}
\label{tab:dataset}
\end{table}

\section{Analysis of Citation Behavior}
We analyze the data sets in various ways to investigate the claim by Verstak et al.~\cite{DBLP:journals/corr/VerstakASHILS14}. First of all, we show that if we divide the cited papers into two time-zones -- those having a time difference of $\leq t$ years with the citer paper and all others, we obtain similar results as described in ~\cite{DBLP:journals/corr/VerstakASHILS14}. However, if we fix the set of old papers, our results suggest that the number of citations that these papers get over the years are indeed affected by the aging behavior.

\subsection{Fraction of citations to (all) `old' papers}
\begin{figure}[!thb]
\[\arraycolsep=.5pt\def\arraystretch{.8}
\begin{array}{cc}
\resizebox{.5\linewidth}{!}{\includegraphics{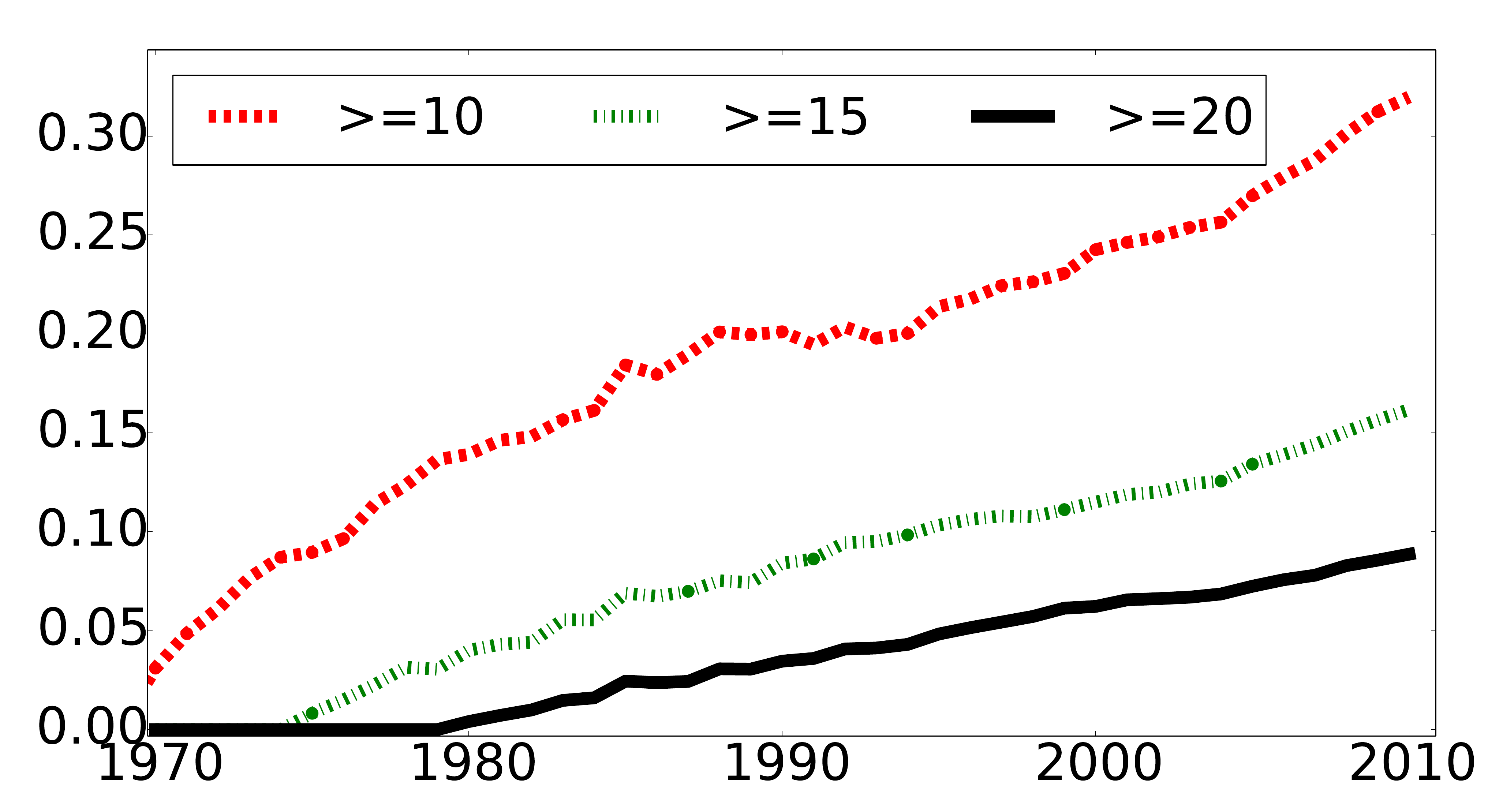}} & \resizebox{.5\linewidth}{!}{\includegraphics{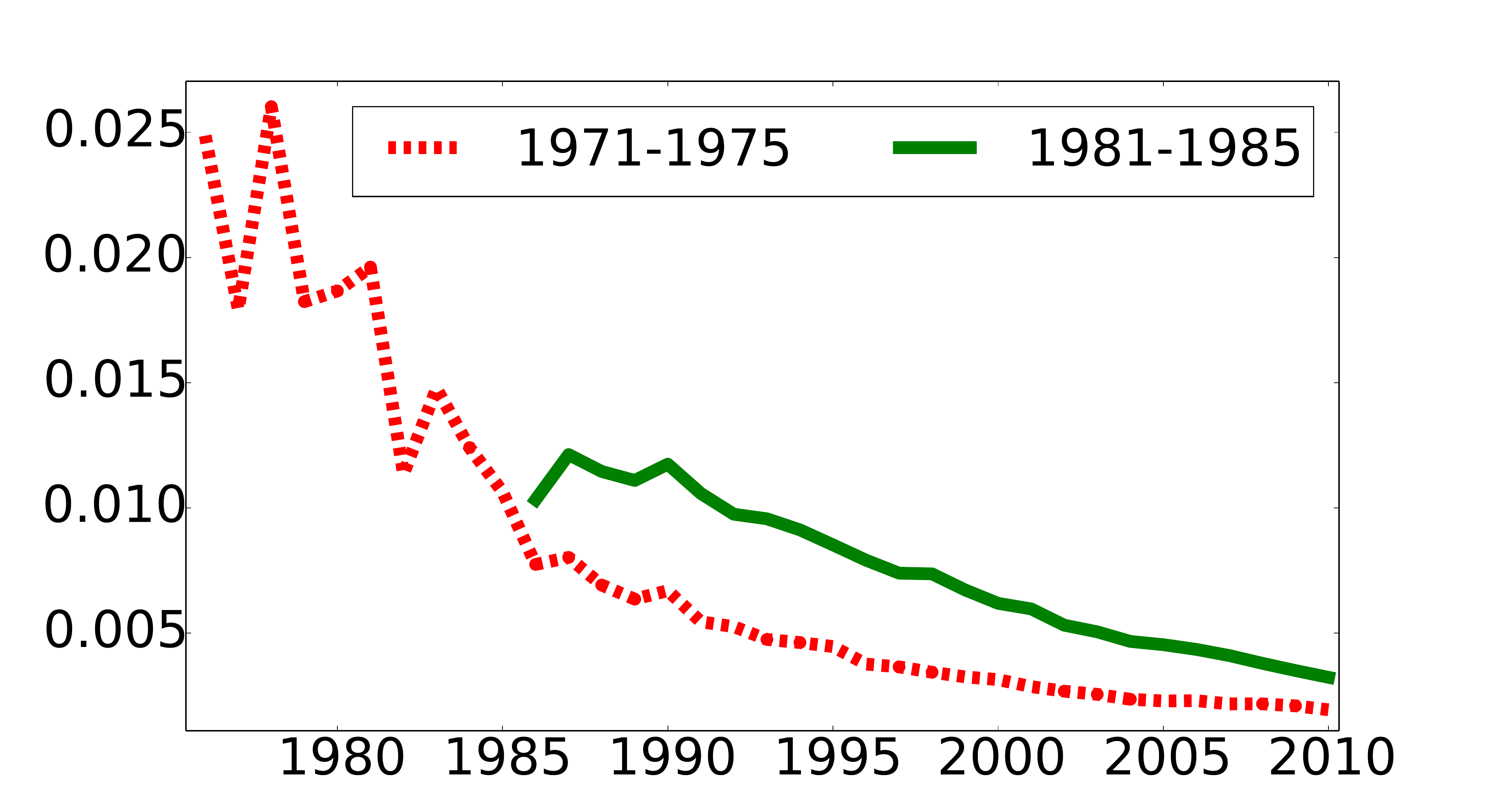}}\\
 (a) & (b)\\ 
\resizebox{.5\linewidth}{!}{\includegraphics{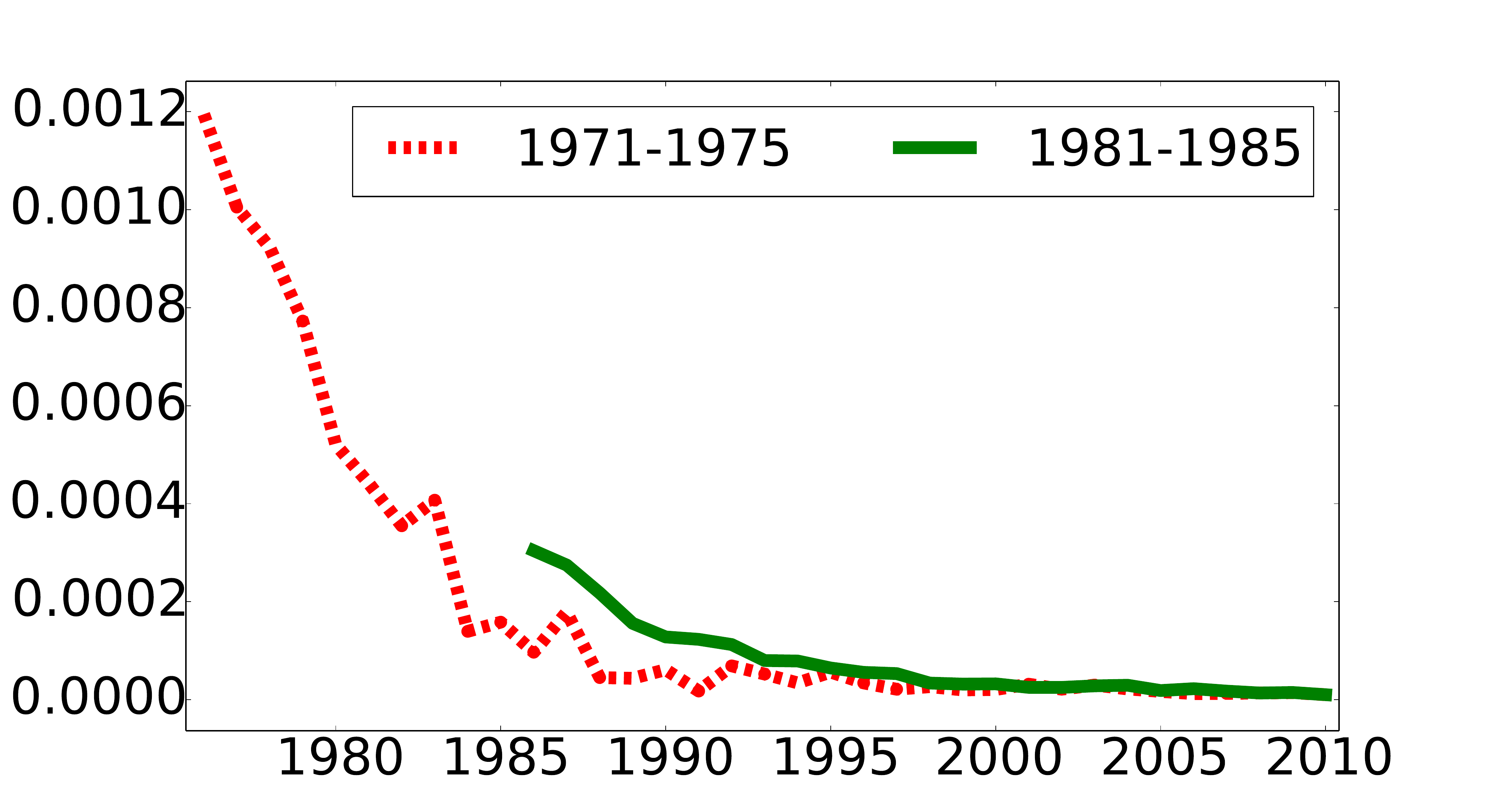}} & \resizebox{.5\linewidth}{!}{\includegraphics{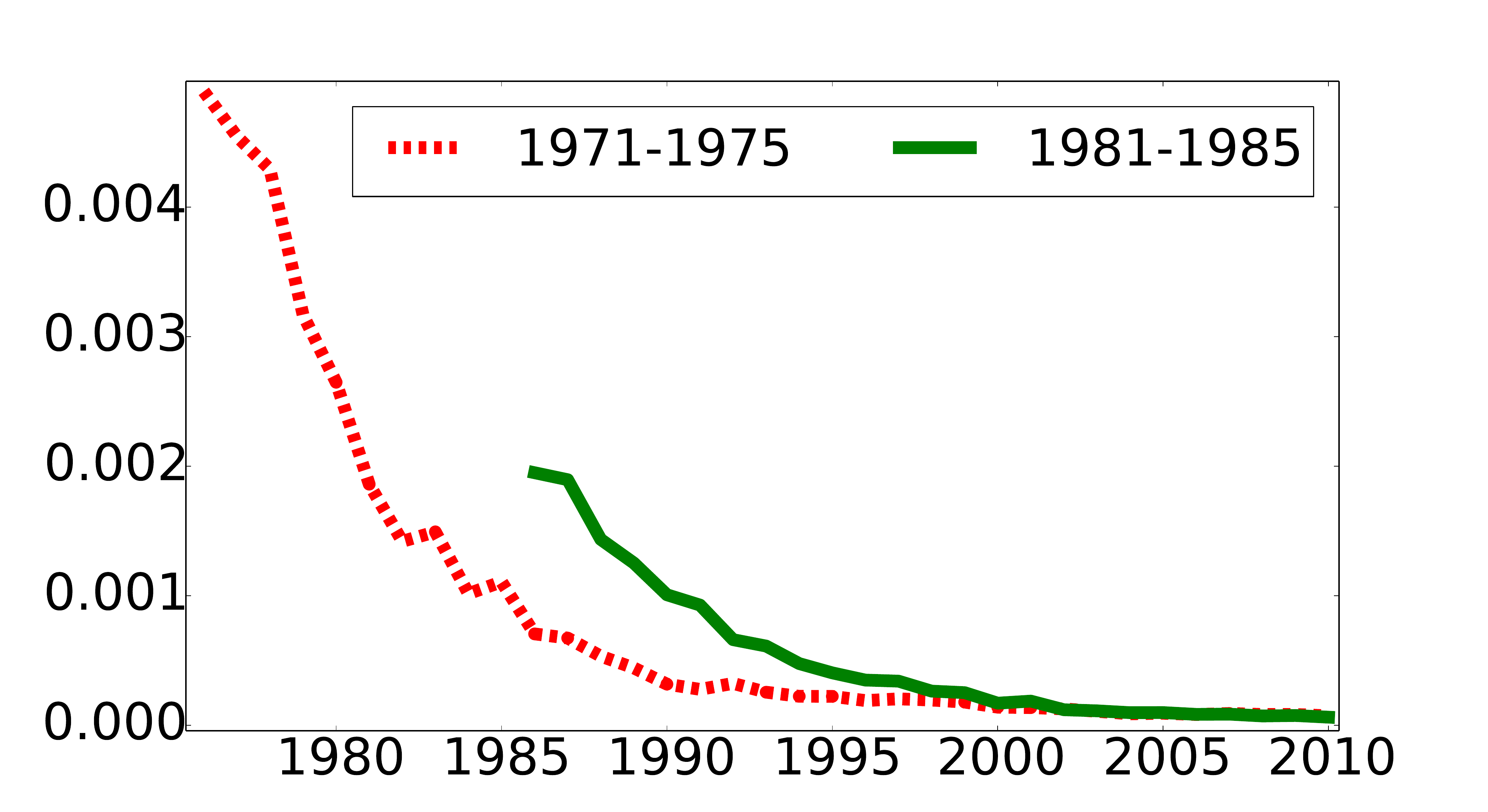}}\\
\vspace{-.4cm} (c) & (d)\\ 
\end{array}
\]
\caption{Fraction of citations between (a) 1970-2010 to all the older papers in three different time-windows ($\geq10$ years, $\geq15$ years and $\geq20$ years, (b) 1975-2010 to a fixed set of 100 most cited papers from 1971-1975 and 1981-1985, (c) 1975-2010 to a fixed set of 100 random papers from 1971-1975 and 1981-1985 and (d) 1975-2010 to a fixed set of 500 random papers from 1971-1975 and 1981-1985.}
\label{fig:citations}
\end{figure}

For the papers published between 1970-2010, Figure \ref{fig:citations}(a) shows fraction of out-citations given to all the papers older than $t$ years for three different values of $t$ ($10,15,20$). For a particular year $y$ and time-window $t$, we compute fraction of older citations $R_{t,y}$ as
\begin{equation}
R_{t,y}= \frac {C_t}{C_y} 
\end{equation}
where $C_y$ represents number of citations from papers published in year $y$ and $C_t$ represents number of citations to papers with publication time difference $\geq t$ from the year $y$. We observe that this is consistent with the claim by Verstak et al.~\cite{DBLP:journals/corr/VerstakASHILS14}, fraction of citations to the older papers is increasing over the years for all the values of $t$. 

Our central claim is that while the fraction of citations to all the older papers might be increasing over time, the set of older papers which are getting more citations is not the same. The above formulation fails to capture the notion that in each successive year, an increasing number of publication set is added to time-window $\geq t$. Thus, the increase in the fraction of citations might just be a byproduct of the increase in the number of publications. We suggest that if instead, the number of publications are fixed for various time-windows and we observe the fraction of citations going to these papers, the results give a very different picture.

As a first experiment, we took 100 most cited articles from the years 1971-1975 and 1981-1985 each. Figure \ref{fig:citations}(b) presents fraction of citations going to each of these sets over the years (i.e., what fraction of the out-going citations in that year went to these set of papers). We clearly see that this fraction is decreasing over the time. Further, we conducted experiments for random set of papers for each year intervals.  Figure \ref{fig:citations}(c and d) show similar observations for 100 and 500 random papers respectively. This observation motivates us to conduct detailed experiments by fixing papers in 10-year buckets and studying the citations they receive over time. 

\subsection{Fraction of citations in 10-year buckets}
\begin{figure}[!thb]
\[\arraycolsep=.5pt\def\arraystretch{.8}
\begin{array}{cc}\vspace{-.3cm}
\resizebox{.5\linewidth}{!}{\includegraphics{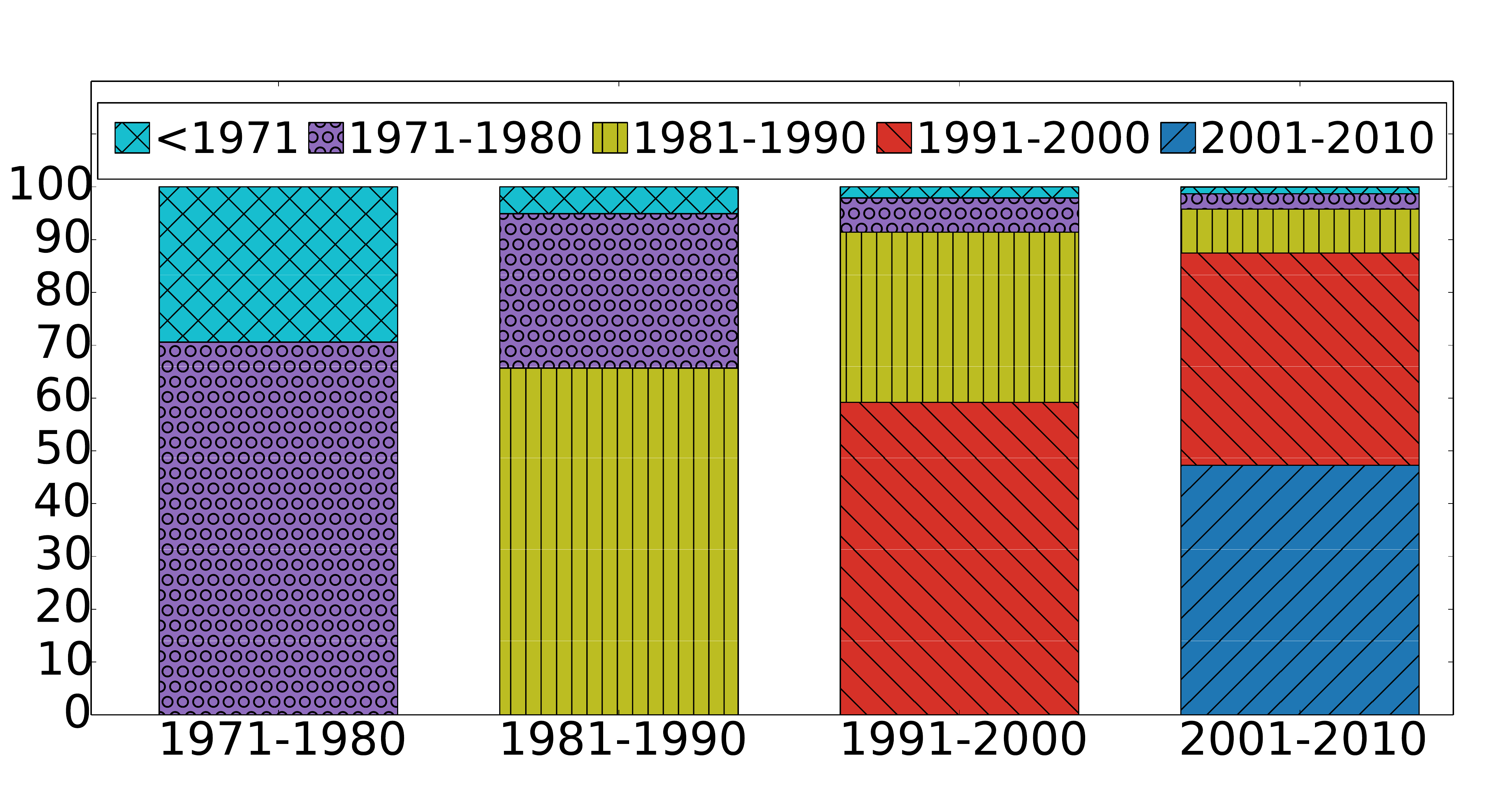}} & \resizebox{.5\linewidth}{!}{\includegraphics{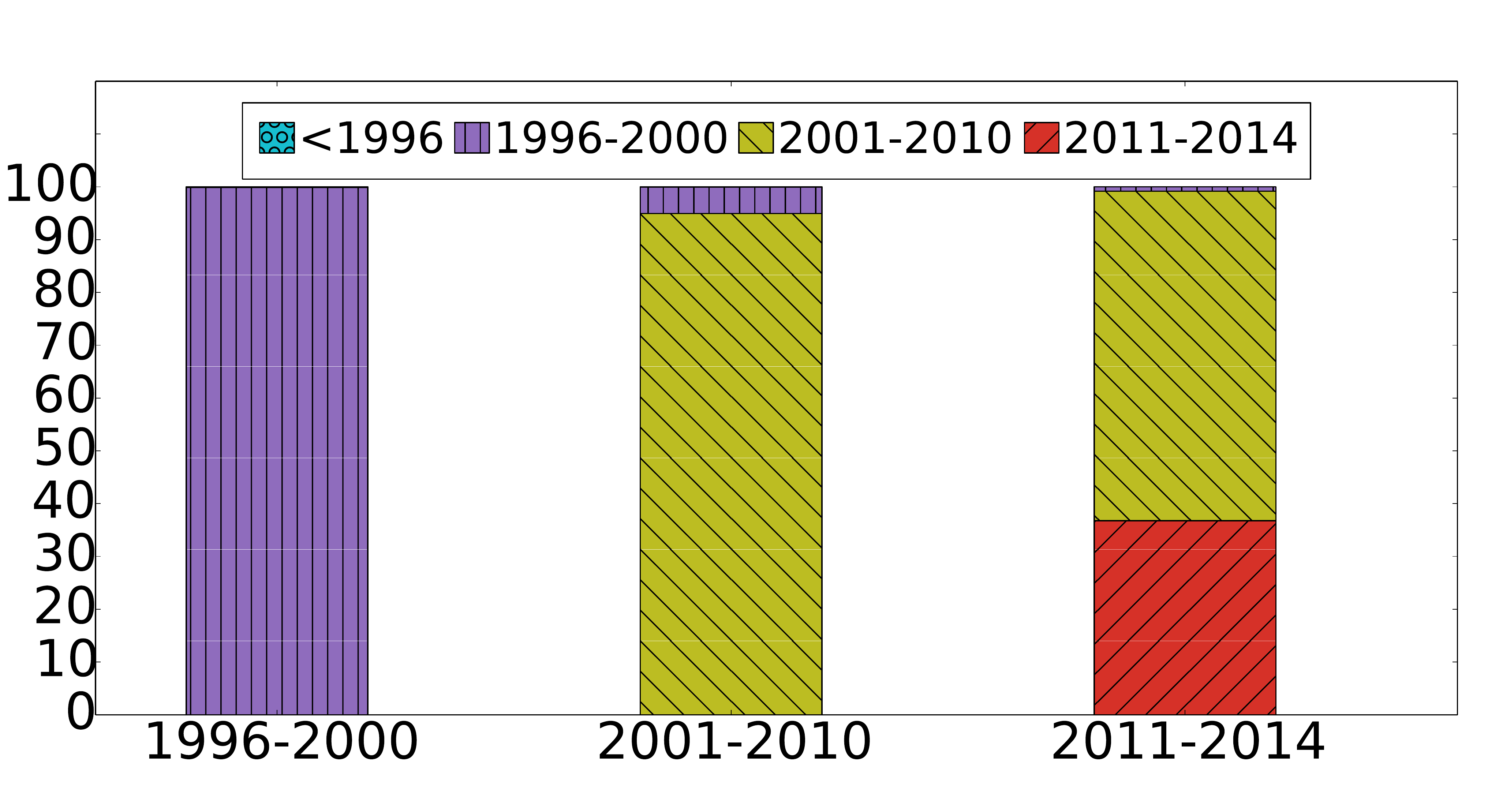}}\\
\vspace{-.4cm} (a) & (b)\\ 
\end{array}
\]
\caption{Stack plot showing citation distribution in 10-year buckets for papers in (a) Computer Science and (b) Biomedical domain,  Each color/texture represents citations made to papers written in a time window, from future papers.  Note that each such slab shrinks dramatically as time passes.}
\label{fig:buckets_all}
\end{figure}

We now group all the papers in the Computer Science domain in different buckets as per publication years, with each bucket consisting of papers published in one decade. In Figure \ref{fig:buckets_all} (a), for each of these buckets, we plot the fraction of citations going to the current bucket and all the previous buckets. We note the following:
\begin{itemize}[noitemsep]
\item The fraction of citations to the same bucket decreases over time (and those to all the older papers increase over time), consistent with the previous observations.
\item If we consider the papers in a given bucket, the citations it receives decreases over the years. For instance, the papers in 1971-1980 received 70.5\% of the citations in that decade but this number reduces to $29.2,6.4,2.8$ in the consecutive decades. This observation holds true for all the consecutive buckets as well.
\end{itemize}

To verify that these observations are universal, we take another dataset from the biomedical domain. We see similar observations with this dataset as well Figure \ref{fig:buckets_all}(b).

\subsection{The case of different Computer Science fields}
\begin{figure*}[!thb]
\[\arraycolsep=.5pt\def\arraystretch{.5}
\begin{array}{llll}\vspace{-.1cm}
\resizebox{.5\linewidth}{!}{\includegraphics{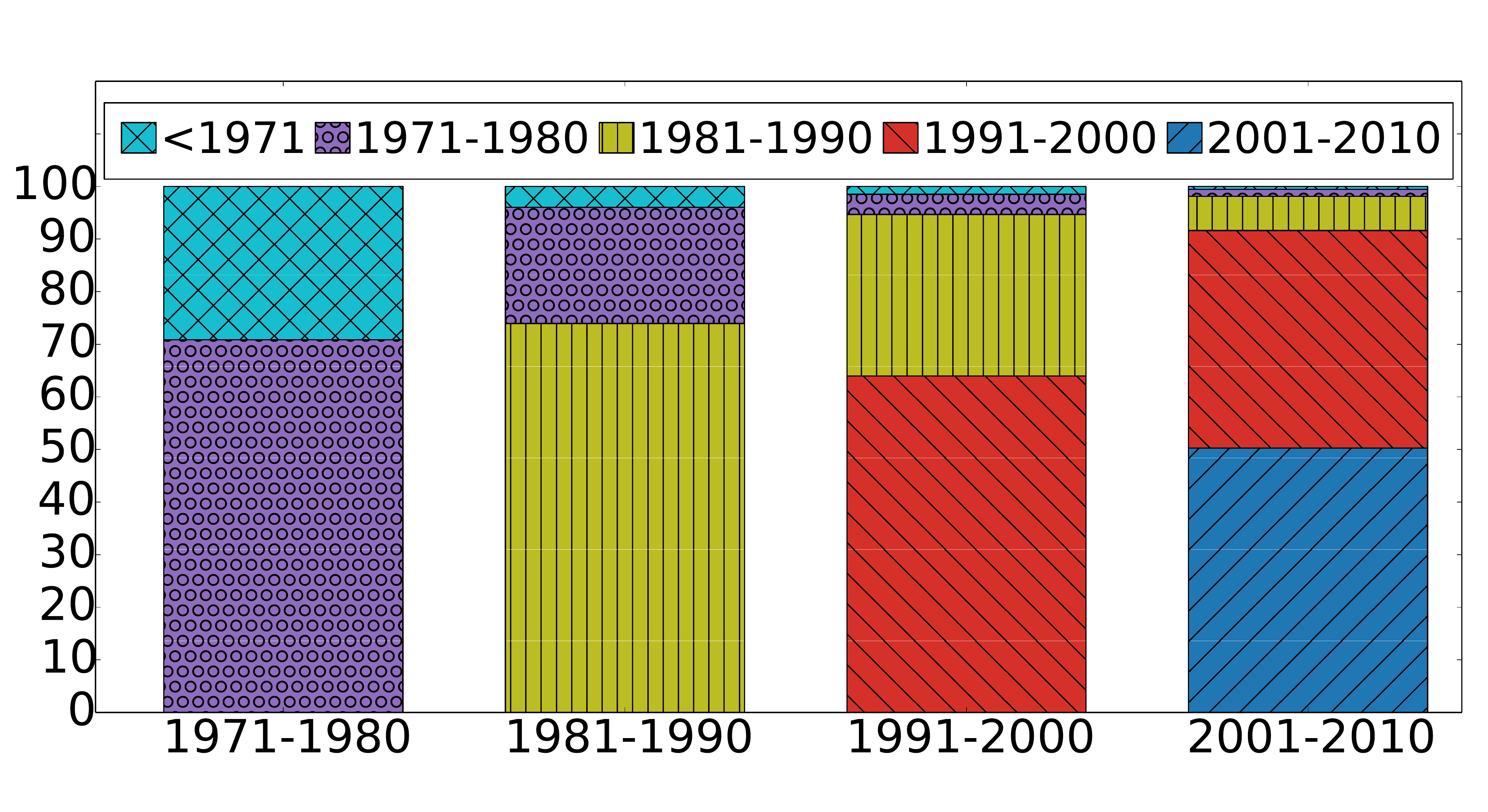}} \resizebox{.5\linewidth}{!}{\includegraphics{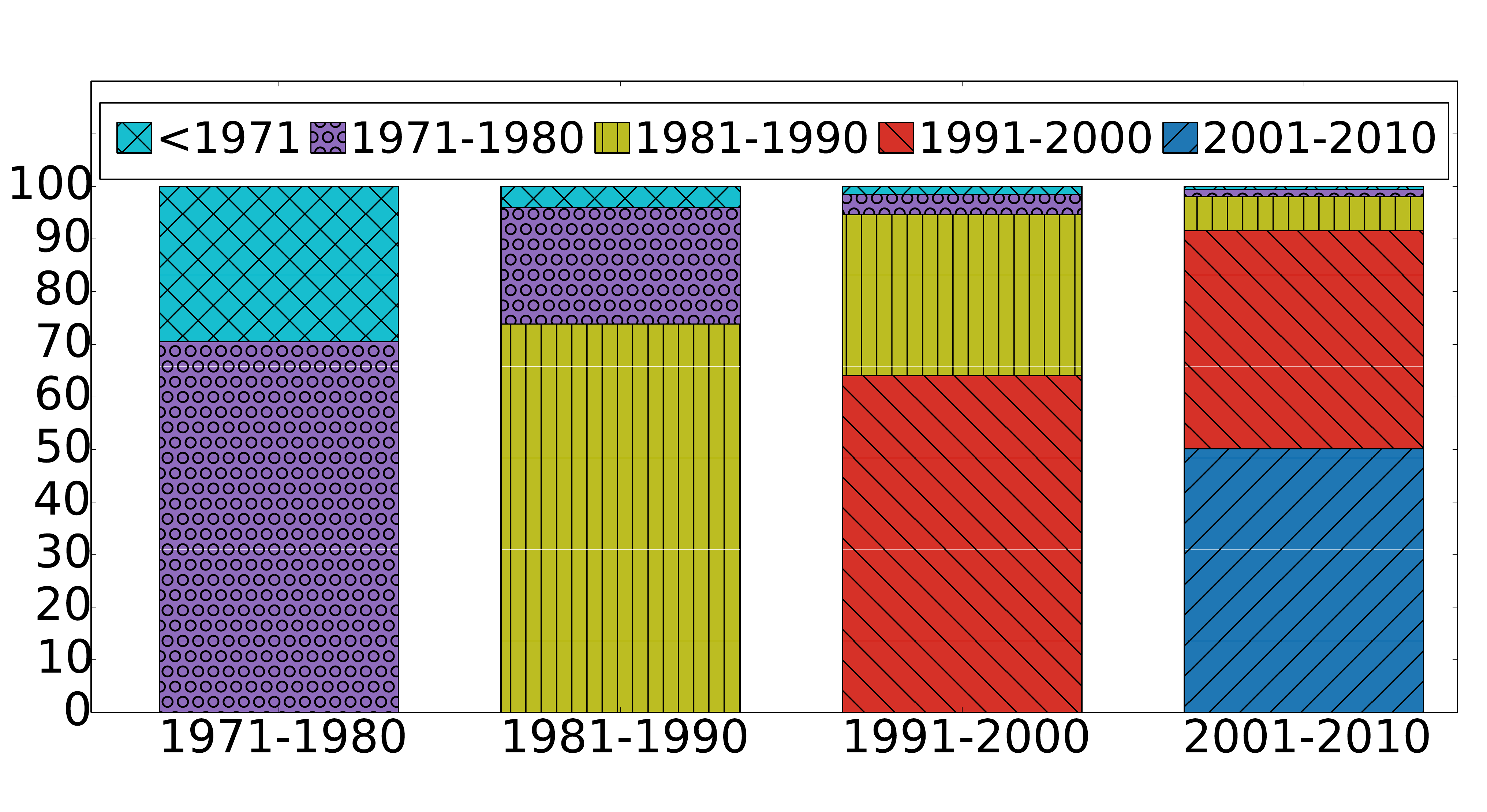}} \\
\hspace{2cm} (a) Decay = 3.32   \hspace{5cm} (b) Decay = 3.31 \\
\resizebox{.5\linewidth}{!}{\includegraphics{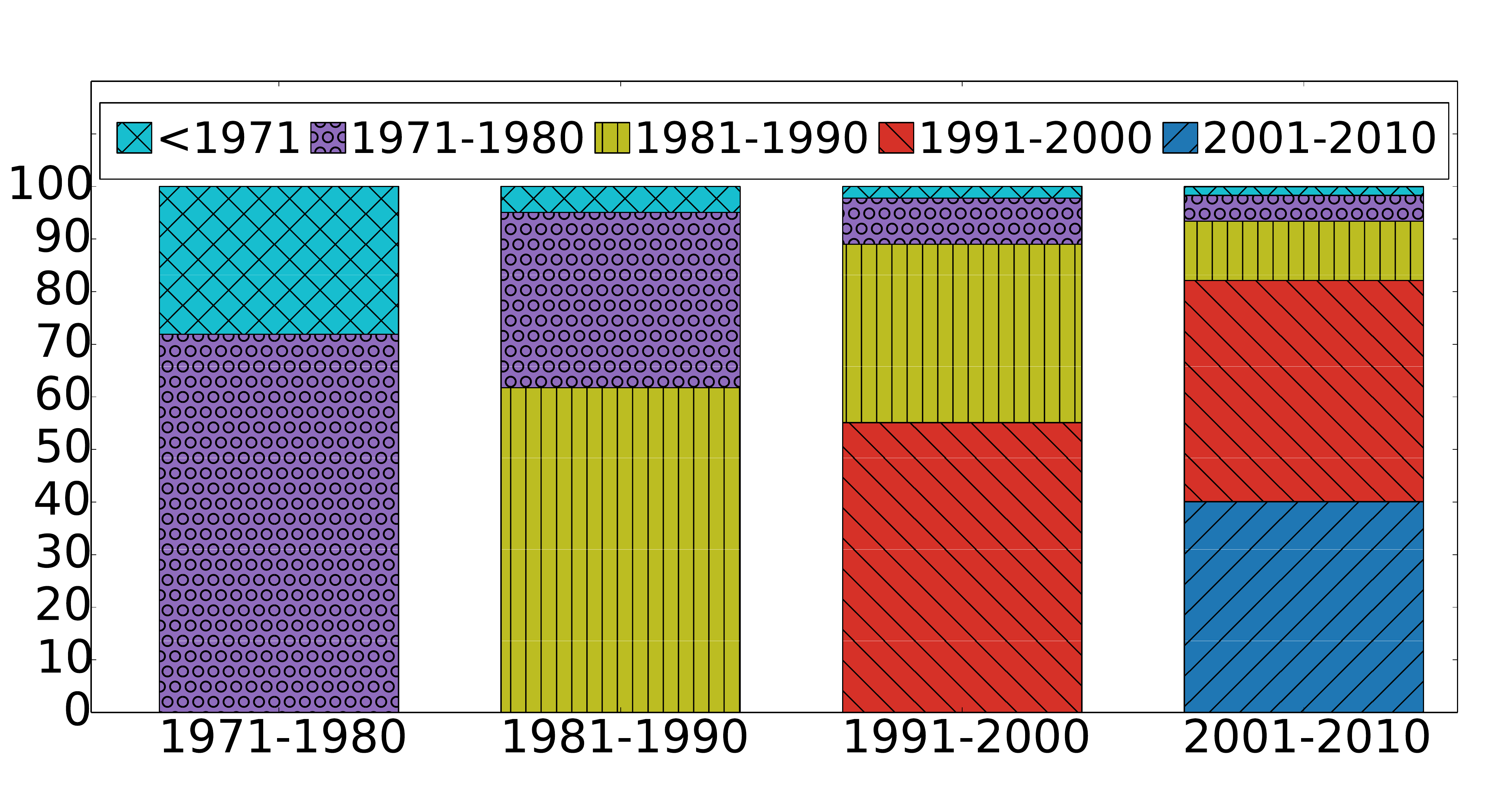}} \resizebox{.5\linewidth}{!}{\includegraphics{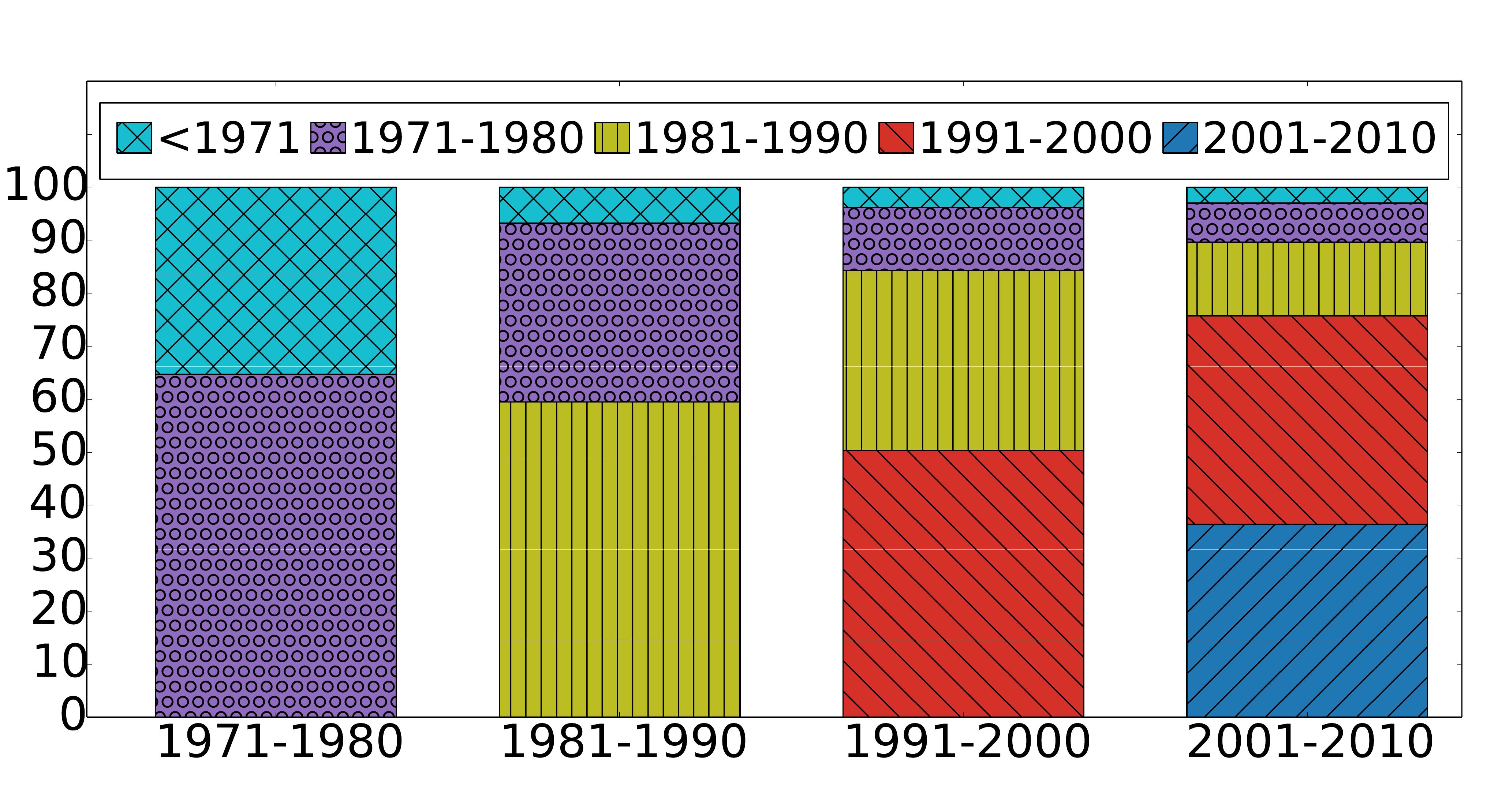}}\\
\hspace{2cm} (c) Decay = 2.29 \hspace{5cm}  (d) Decay = 2.02 \\ 
\end{array}
\]
\caption{Stack plot showing citation distribution in 10-year buckets for (a) Data Mining, Machine Learning, Artificial Intelligence, Natural Language processing and Information Retrieval, (b) Distributed and Parallel Computing, Hardware and Architecture and  real time and Embedded Systems, (c) Algorithms and Theory, Programming Languages and Software Engineering and (d) Algorithms,}
\label{fig:fields}
\end{figure*}

\begin{figure*}[!thb]
\[\arraycolsep=.5pt\def\arraystretch{.5}
\begin{array}{llll}
\resizebox{.5\linewidth}{!}{\includegraphics{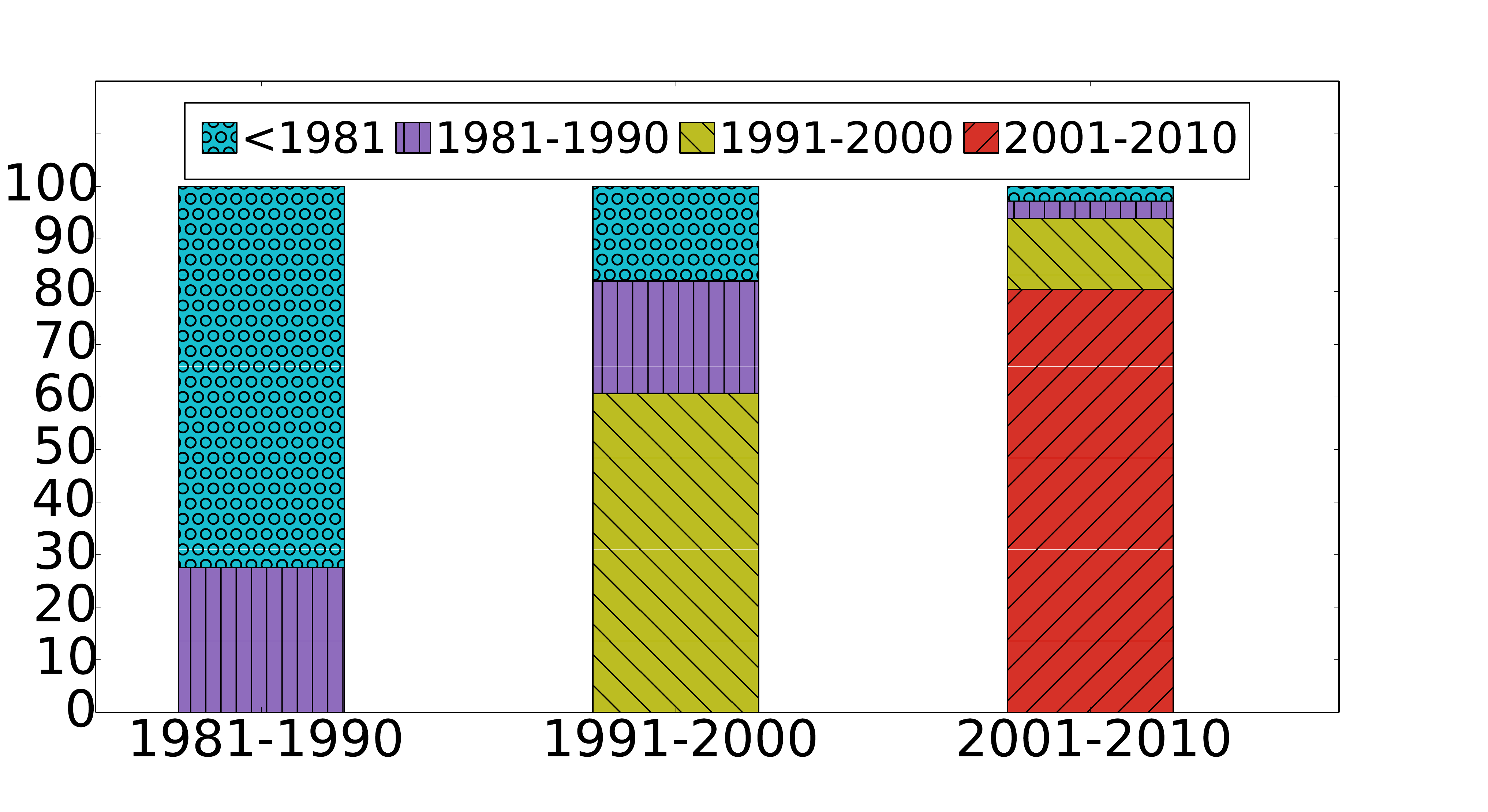}}  \resizebox{.5\linewidth}{!}{\includegraphics{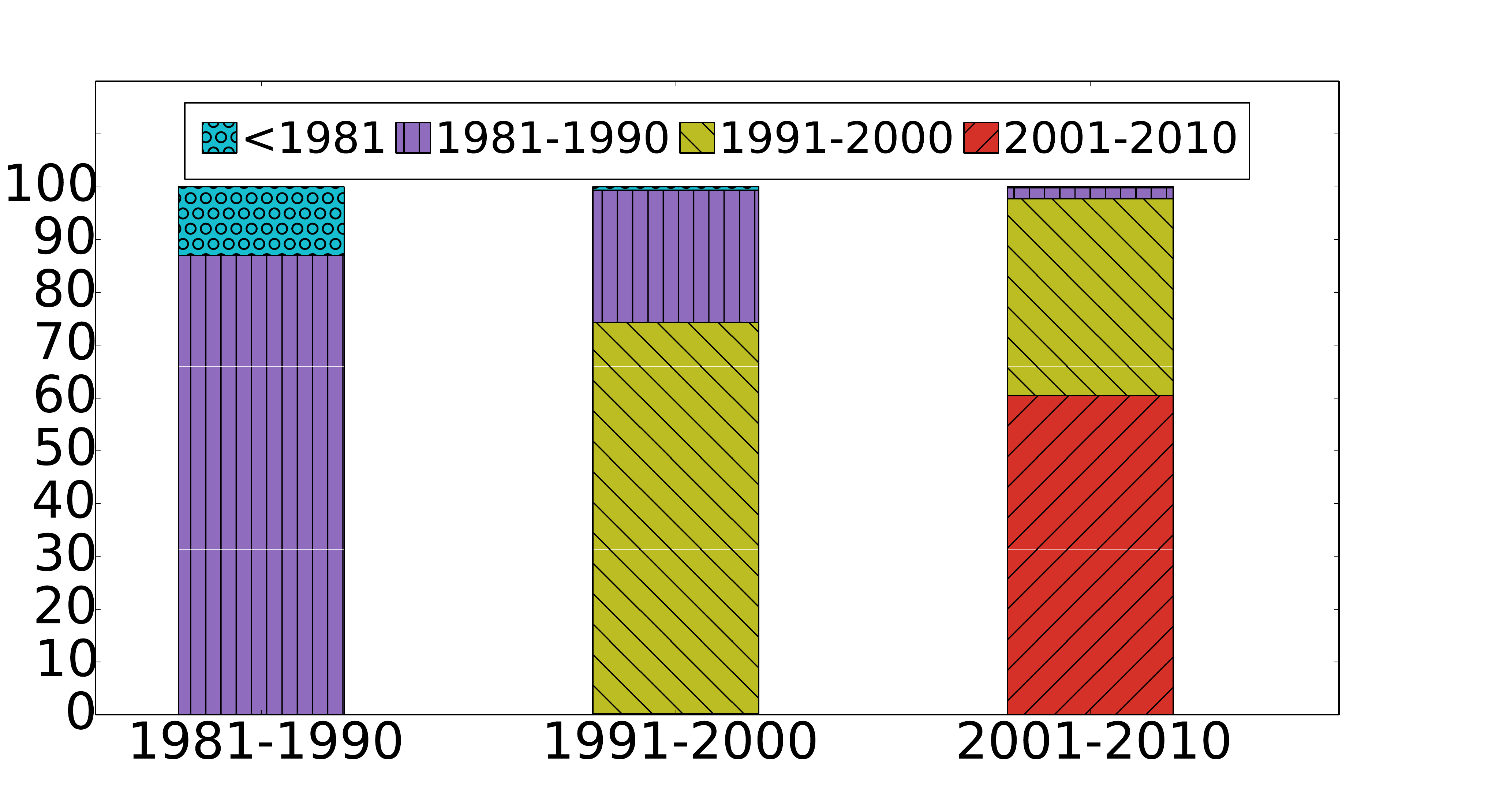}}\\
\hspace{2cm} (a) Decay= 3.97  \hspace{5cm}   (b)Decay= 5.61 \hspace{2.1cm} \\
\resizebox{.5\linewidth}{!}{\includegraphics{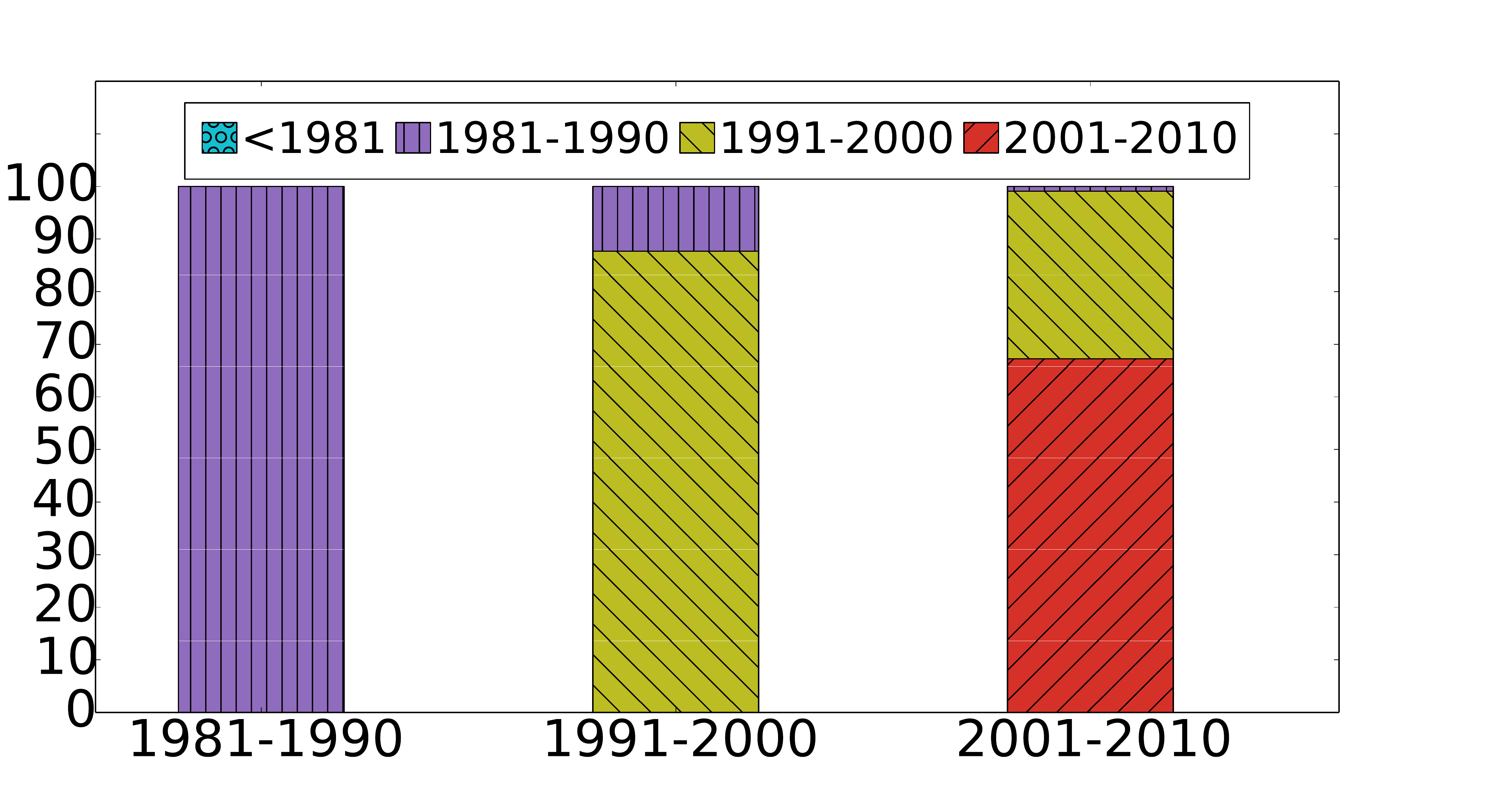}}  \resizebox{.5\linewidth}{!}{\includegraphics{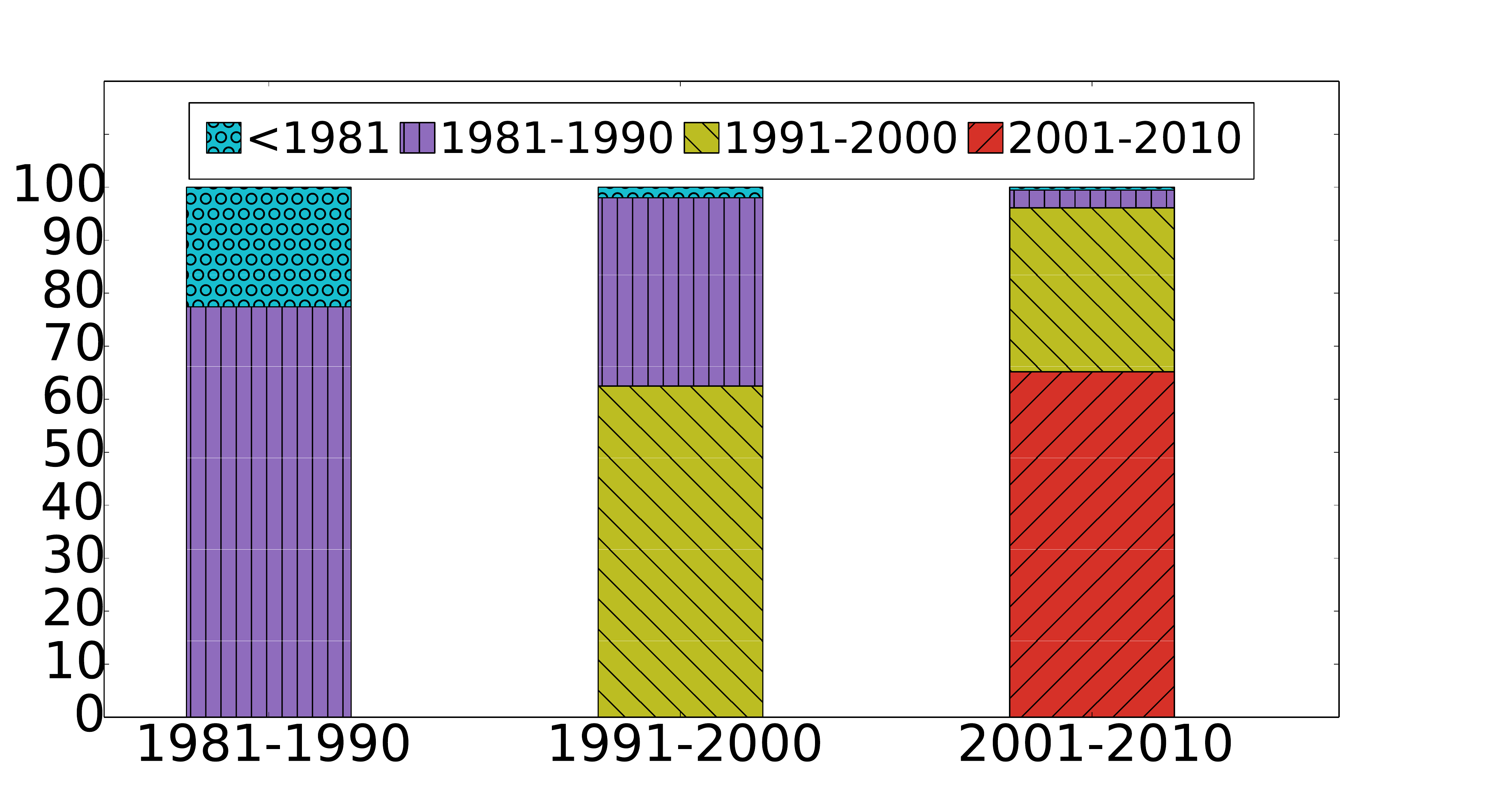}}\\
\hspace{2cm}(c)Decay= 6.74 \hspace{5cm}(d)Decay= 4.52\\ 
\end{array}
\]
\caption{Stack plot showing citation distribution in 10-year buckets for (a) SIGMOD, (b) SIGIR, (c) ICML-NIPS and (d) VLDB-ICDE}
\label{fig:confs}
\end{figure*}

We now repeat the same experiment by plotting the incoming citations to various sub-fields of computer science domain for various buckets. For instance, in Figure \ref{fig:fields}(a), the stacked plot at 1981-1990 denotes that among all the citations made to Data Mining etc. fields in that decade, what fractions of citations were made to the papers in the 1981-1990 (yellow), 1971-1980 (violet) and 1961-1970 (blue). We plot this for the following combinations of fields:
\begin{itemize}[noitemsep]
\item Data Mining, Machine Learning, Artificial Intelligence, Natural Language processing and Information Retrieval -- Figure \ref{fig:fields}(a)
\item Distributed and Parallel Computing, Hardware and Architecture and  real time and Embedded Systems -- Figure \ref{fig:fields}(b)
\item Algorithms and Theory, Programming Languages and Software Engineering -- Figure \ref{fig:fields}(c)
\end{itemize}

We see that earlier observations hold true for all these stacked bar charts as well, with fraction of citations going to older papers in that field decreasing over time. The amount by which this decay happens, though, is different for different fields (lower in case of Algorithms and Theory than others). If we study only the Algorithms field, the decay is even slower (Figure \ref{fig:fields}(d)).  

To quantify this decay, we define a decay metric that measures the decrease in fraction of older citations for a field over the years. For papers published in a given time-interval, we compute the ratio of citations in consecutive year buckets and take a geometric mean over all such ratios. A high value of this geometric mean indicates a higher decay. So, in Figure \ref{fig:fields}, Algorithms and Theory field has the smallest decay. If we consider only the Algorithms field, decay is even smaller.

\subsection{The case of different Computer Science Conferences}
We further perform this analysis for various conferences in the field of Computer Science. Thus, for various time-points, we plot that among all the citations going to that particular conference, how many citations were made to the same time-point, previous one and so on. This analysis is performed for SIGMOD (Figure \ref{fig:confs}(a)), SIGIR (Figure \ref{fig:confs}(b)), ICML-NIPS (Figure \ref{fig:confs}(c)) and VLDB-ICDE (Figure \ref{fig:confs}(d)). The decay factors obtained for these conferences is much higher than observed for various fields in Figure \ref{fig:fields}. As a side remark, we observe that for SIGMOD, the fraction of citations going to all the older papers decreases over time contrary to that reported in Verstak \emph{et al.} \cite{DBLP:journals/corr/VerstakASHILS14}. However, our observation regarding the aging effect still holds true.

Further, we attempt to correlate the decay factor with the average value of 10 year impact factor~\cite{garfield2006history} of the conferences in Table \ref{tab:decay_impact_factor} and we find that they are negatively correlated, with high decay factor implying smaller Impact Factor.
  
\begin{table}[!thb]
 \centering
 \caption{Correlation between decay and average value of 10-year impact factor.}\label{decay_impact_factor}
  \begin{tabular}{|c|c|c|}
  \hline
  \textbf{Conf. Name}&\textbf{Decay Factor}&\textbf{Avg. IF$_{10}$}   \\\hline
  SIGMOD&3.97&3.50\\\hline
  SIGIR&5.61&2.77\\\hline
  ICML-NIPS&6.74&1.84\\\hline
  VLDB-ICDE&4.52&2.79\\\hline
 \end{tabular}
\label{tab:decay_impact_factor}
\end{table}

\section{Discussions}
A thorough analysis by fixing the set of cited papers in 10-year buckets helps us to reconcile the two contradictory views of the aggregate fate of ancient papers, that is, while the fraction of citations to all the papers older than a fixed number of years increases over time, that to a fixed set of old papers tends to decrease over time. 
\bibliographystyle{abbrv}
\bibliography{main.bib}

\end{document}